# Temporal dark polariton solitons

Yaroslav V. Kartashov[1,2,*] and Dmitry V. Skryabin[3,4]

[1]ICFO-Institut de Ciencies Fotoniques, The Barcelona Institute of Science and Technology, 08860 Castelldefels (Barcelona), Spain
[2]Institute of Spectroscopy, Russian Academy of Sciences, Troitsk, Moscow, 142190, Russian Federation
[3]Department of Physics, University of Bath, BA2 7AY, Bath, United Kingdom
[4]ITMO University, Kronverksky Avenue 49, St. Petersburg 197101, Russian Federation

**We predict that strong coupling between waveguide photons and excitons of quantum well embedded into waveguide results in the formation of hybrid dark and anti-dark light-matter solitons. Such temporal solitons exist due to interplay between repulsive excitonic nonlinearity and giant group velocity dispersion arising in the vicinity of excitonic resonance. Such fully conservative states do not require external pumping to counteract losses and form continuous families parameterized by the power-dependent phase shift and velocity of their motion. Dark solitons are stable in the considerable part of their existence domain, while anti-dark solitons are always unstable. Both families exist *outside* forbidden frequency gap of the linear system.**

Using strong light-matter coupling and record high levels of nonlinearity in the exciton-polariton micro-resonators, waveguides, and periodic lattices is becoming a rapidly developing branch of photonics showing a plethora of fundamental physical effects at the interface between optics and condensed matter physics [1]. Recently planar waveguides and photonic wires operating in the regime of strong coupling between photons and quantum-well excitons have been demonstrated [2-5] and used for observation of temporal and spatio-temporal solitons. The effective nonlinear parameter $\gamma$ [6] in these waveguides can reach $10^5$ W$^{-1}$m$^{-1}$, that exceeds highest values of nonlinear parameters demonstrated in optical fibers and semiconductor waveguides by several orders of magnitude [7]. Thus, such waveguides represent an excellent experimental platform for the demonstration of ultra-low-power self-sustained excitations.

Optical modes of a waveguide with single or multiple quantum wells can couple to an excitonic resonance [2-4]. In the presence of waveguide spatial localization is mediated by the total internal reflection, i.e. no Bragg mirrors in the cladding are necessary. Under these conditions the photon component of the polariton quasi-particle is given by the electro-magnetic waveguide mode detuned far away from the waveguide cut-off and having the dispersion profile linear over the bandwidth of several excitonic resonances. The latter are well described by the Lorentz oscillator model with the resonance energy upshifted through the repulsive exciton-exciton interaction, which produces net defocusing nonlinearity for polar-itons. This system features upper and lower polariton branches having respectively anomalous and normal group velocity dispersions. The latter case in combination with the defocusing excitonic nonlinearity gives rise to *bright temporal solitons* that have been demonstrated experimentally in [2], while the former case suggests the existence of *dark temporal solitons* that are under investigation in this Letter.

One should mention the link between *temporal* excitations studied in this Letter and *spatial* dark polariton solitons forming in planar micro-resonators upon interaction with an obstacle [8,9]. As these solitons propagate with some slow velocities of few percent of the light velocity, they also can be considered as spatio-temporal structures, but in the case when dispersion of the light mode is shaped by the waveguide cut-off equivalent to the cavity resonance. Solitons of the Maxwell-Lorentz systems have been studied in [10,11], but outside the context of recent and ongoing exciton-polariton experiments. Note also certain similarity between mechanism of soliton formation in our system and the effect of self-induced transparency in the exciton medium introduced in [12,13].

Here we use Maxwell-Lorentz system of coupled dimensionless equations for amplitudes of the photonic guided mode $A_\text{p}$, confined in the direction perpendicular to propagation direction $z$, and excitonic polarization $A_\text{e}$ to describe evolution of nonlinear temporal excitations in a semiconductor waveguide with embedded quantum wells:

$$2i\beta_e v_\text{g}^{-1}(v_\text{g}\partial/\partial z + \partial/\partial t + \nu_\text{p})A_\text{p} = -k_\text{e}^2 A_\text{e},$$
$$-2i(\partial/\partial t + \nu_\text{e})A_\text{e} = \kappa A_\text{p} - g A_\text{e} |A_\text{e}|^2. \quad (1)$$

Here and below we are accounting for the dispersion induced by the excitonic resonance, for the defocusing nonlinearity induced by the exciton-exciton repulsion, photonic losses and the exciton decoherence rate. We further rewrite Eqs. (1) in dimensionless units:

$$i\frac{\partial \psi_\text{p}}{\partial \tau} = -i(1-v)\frac{\partial \psi_\text{p}}{\partial \xi} - \Omega\psi_\text{e} - i\alpha_\text{p}\psi_\text{p},$$
$$i\frac{\partial \psi_\text{e}}{\partial \tau} = iv\frac{\partial \psi_\text{e}}{\partial \xi} - \Omega\psi_\text{p} + |\psi_\text{e}|^2\psi_\text{e} - i\alpha_\text{e}\psi_\text{e}. \quad (2)$$

Here, $\tau = t/t_0$ is the dimensionless time, where $t_0$ is the characteristic pulse duration; $\xi = (z - tv_\text{s})/t_0 v_\text{g}$ is the dimensionless distance along the waveguide; $v_\text{s}$ is the soliton velocity and $v_\text{g}$ is the group velocity for photonic mode at the exciton frequency $\omega_\text{e}$; $v = v_\text{s}/v_\text{g}$; the coupling strength between photons and excitons is described by the dimensionless parameter $\Omega = (k_e^2 v_g t_0^2 \kappa / 4\beta_e)^{1/2}$, where $k_e = \omega_e/c$ is the wavenumber, $\beta_e$ is the propagation constant of the photonic guided mode, $\kappa$ is the actual light-matter coupling rate, $\alpha_{p,e} = \nu_{p,e}\tau_0$ characterize losses. Dimensionless amplitudes $\psi_{p,e}$ are connected with $A_{p,e}$ by the relations $\psi_p = (gt_0\beta_e\kappa/2k_e^2 v_g)^{1/2} A_p$ and $\psi_e = (gt_0/2)^{1/2} A_e$, where $g$ is the strength of exciton-exciton interactions (see [2] for connection between $g$ and $\gamma$). We are selecting $t_0 = (4\beta_e/\kappa k_e^2 v_g)^{1/2}$ to make $\Omega = 1$. Further we treat time $\tau$ as an evolution variable and search for nonlinear excitations localized in $\xi$. Characteristic temporal widths of cores of dark solitons considered here are expected to be close to typical duration of bright polariton solitons observed in [2] that ranged from hundreds of fs to few ps and carried energy of hundreds of fJ. For $\tau_0 \sim 350$ fs one has $\alpha_p \approx 0.0236$, $\alpha_e \approx 0.004$ [2] and we initially omit these small losses when searching for soliton solutions.

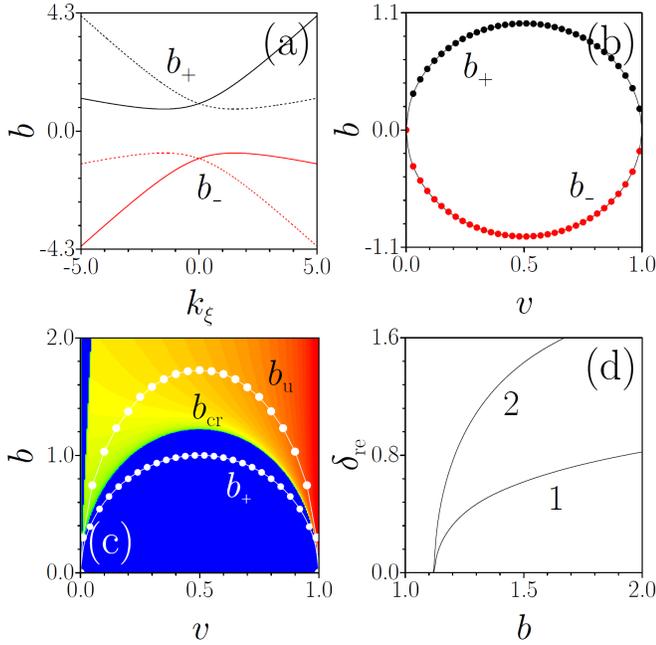

Fig. 1. (Color online) (a) Propagation constant of linear wave versus wavenumber $k_\xi$ at $v = 0.2$ (solid lines) and $v = 0.8$ (dashed lines). Black (red) curves show upper (lower) branches of the dispersion relation. (b) Upper and lower edges of the gap versus $v$. (c) Modulation instability growth rate $\delta_{re}$ for background plane waves vs $(v,b)$. Logarithmic scale is used for different $\delta_{re}$ levels to stress stability border located at $b = b_{cr}$. Lines with circles indicate upper border of the gap $b_+$ and upper border of the existence domain for dark solitons $b_u$. (d) $\delta_{re}(b)$ dependencies for $v = 0.3$ (curve 1) and $v = 0.7$ (curve 2). Here and below $\Omega = 1$.

Looking for solutions of the linearized Eq. (2) in the form $\psi_{p,e} = a_{p,e} \exp(ik_\xi \xi - ib\tau)$, where $b$ is the frequency detuning and $k_\xi$ is the wavenumber shift, we find the dispersion relation with two branches

$$b_\pm = -\frac{1}{2}[-k_\xi(1-2v) \pm (k_\xi^2 + 4\Omega^2)^{1/2}] \quad (3)$$

shown in Fig. 1(a). As one can see the branches are separated by the gap, which closes at $v = 0$ and $v = 1$ and has maximal width at $v = 0.5$, see Fig. 1(b). It was shown in [2] that in the presence of nonlinearity bright solitons exist within the gap and emerge from its lower edge (lower polariton branch), where group velocity dispersion is normal

$\partial^2 b_-/\partial k_\xi^2 > 0$. Bright soliton width greatly increases at both gap edges, so that the best localization is achieved in the middle of the gap.

Here we are interested in *dark solitons* that are expected to exist close to the upper polariton branch, where group velocity dispersion is anomalous (let's recall that nonlinearity is defocusing). We look for the soliton solutions in the form $\psi_{p,e}(\xi,\tau) = a_{p,e}(\xi)\exp[i\alpha\xi + i\phi_{p,e}(\xi) - ib\tau]$, where $\alpha$ is the soliton wavenumber shift, $a_{p,e}, \phi_{p,e}$ are the amplitudes and phases of two components that approach constant asymptotic values at $\xi \to \pm\infty$. Substitution of this expression into (2) and separation of real and imaginary parts yields the system

$$\begin{aligned}
-(1-v)(\alpha + \partial\phi_p/\partial\xi)a_p &= -ba_p - \Omega a_e \cos(\phi_e - \phi_p), \\
(1-v)\partial a_p/\partial\xi &= -\Omega a_e \sin(\phi_e - \phi_p), \\
-v(\alpha + \partial\phi_e/\partial\xi)a_e &= \Omega a_p \cos(\phi_e - \phi_p) - a_e^3 + ba_e, \\
v\partial a_e/\partial\xi &= -\Omega a_p \sin(\phi_e - \phi_p).
\end{aligned} \quad (4)$$

Using methodology similar to that developed for analysis of solitons in fiber Bragg gratings [14], one can show that Eqs. (4) have two integrals of motion:

$$\begin{aligned}
(1-v)a_p^2 - va_e^2 &= C_1, \\
-b(a_p^2 + a_e^2) + a_e^4/2 - 2\Omega a_p a_e \cos(\phi_e - \phi_p) &= C_2,
\end{aligned} \quad (5)$$

where constants $C_{1,2}$ can be determined from the limits of the desired solution at $\xi \to \infty$. Dark solitons, which we study here, reside on the plane wave background. The analysis of the first of Eqs. (4) at $\xi \to \infty$ shows that soliton's wavenumber shift $\alpha = (2v-1)b/[2v(1-v)]$ is uniquely defined by the frequency detuning $b$ (that plays here the role of propagation constant) and soliton velocity $v$. Eqs. (4) predict that only two asymptotic values of phase difference $(\phi_e - \phi_p)_{\xi \to \infty} = 0, \pm\pi n$ ($n \in \mathbb{N}$) admit the plane wave solutions. Dark solitons can exist only for the $\pm\pi n$ case as follows from analysis of Eq. (6) for amplitude of excitonic component derived below. This sets the background amplitudes to $a_{e\infty}^2 = [b/2(1-v)] - [2v\Omega^2/b]$ and $a_{p\infty} = (2v\Omega/b)a_{e\infty}$, and also imposes a constraint on $b \geq [4v(1-v)\Omega^2]^{1/2}$. This lower limit for $b$ (where both asymptotic amplitudes $a_{p,e}$ become zero) defines cutoff for the existence of dark solitons, which *exactly coincides* with the upper edge $b_+$ of the forbidden gap [Fig. 1(b)]. The condition $b \geq [4v(1-v)\Omega^2]^{1/2}$ implies that dark solitons can bifurcate only from the upper edge of the gap, where linear dispersion relation (3) predicts negative sign of $\partial^2 b_+/\partial k_\xi^2$, corresponding to the anomalous dispersion.

In order to analyze stability of the soliton background we assume $\psi_{p,e} = [a_{p,e\infty} e^{i\phi_{p,e\infty}} + W_{p,e} e^{\delta\tau + ik_\xi\xi} + V_{p,e}^* e^{\delta^*\tau - ik_\xi\xi}]e^{i\alpha\xi - ib\tau}$, where $W_{p,e}, V_{p,e}$ are the amplitudes of small perturbation, $k_\xi$ is the wavenumber, and $\delta = \delta_{re} + i\delta_{im}$ is the increment. We found that the background is stable in the region $b_+ < b < b_{cr}$ and is unstable at $b > b_{cr}$, see color map with $\delta_{re}(b,v)$ dependence in Fig. 1(c). The maximal for all $k_\xi$ instability growth rate is shown in Fig. 1(d) for two selected velocities $v$.

Using the above expressions for $a_{p\infty}$ and $a_{e\infty}$ we define constants $C_1 = -2(1-v)va_{e\infty}^4/b$ and $C_2 = (4v-3)a_{e\infty}^4/2$ in Eqs. (5). They can be used to transform Eqs. (4) into a single equation for the amplitude of excitonic component $a_e(\xi)$ of dark soliton:

$$\frac{1}{2}\left(\frac{\partial a_e}{\partial\xi}\right)^2 = -\frac{1}{8v^2 a_e^2}\left[\frac{(a_e^4 + 3a_{e\infty}^4)^2}{4} - \frac{ba_e^2(a_e^2 - a_{e\infty}^2)^2}{1-v} - 4a_e^2 a_{e\infty}^6\right] \quad (6)$$

Eq. (6) can be used to derive an implicit analytical solution. Once this solution is found, the photonic component and the phase difference can be obtained from Eq. (5). However, it is more convenient to analyze (6) graphically by considering it as an energy conservation law $T + P = 0$ for the effective particle of unit mass [i.e. $T = (1/2)(\partial a_e / \partial \xi)^2$] moving inside the potential, see Fig. 2(a). This potential has one maximum corresponding to the background wave and it also admits two soliton solutions, corresponding to the effective particle moving from the maximum either to the left or to the right, reaching a corresponding turning point, where $P$ value coincides with that in the equilibrium (red dots), and coming back to the equilibrium. The profile of the dark soliton corresponding to the motion from the equilibrium to the left towards smaller amplitudes is shown in Fig. 3(a). The other soliton corresponding to the effective particle motion towards the larger amplitudes, is the anti-dark soliton in the form of the hump, residing on the nonzero background, see Fig. 3(b). It should be stressed that the potential $P$ does not have local maxima for $(\phi_e - \phi_p)_{\xi \to \infty} = 0$, hence this choice of $(\phi_e - \phi_p)_{\xi \to \infty}$ cannot give rise to the localized solutions.

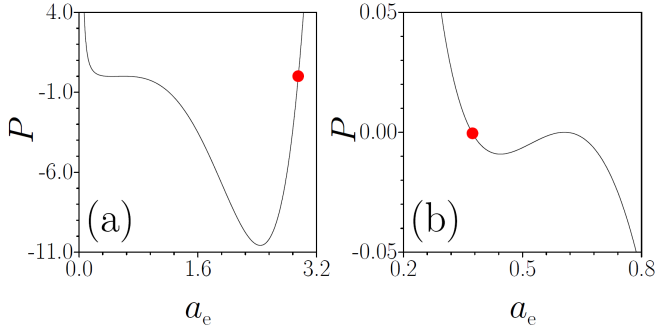

Fig. 2. (Color online) Potential energy for the associated effective particle versus amplitude of excitonic component at $v = 0.5$, $b = 1.2$. Red dots indicate turning points for the anti-dark (a) and dark (b) solitons.

Dark and anti-dark solitons feature qualitatively different phase distributions. In simulations we always set $\phi_p = 0$, $\phi_e = \pi$ at $\xi \to -\infty$. For dark solitons both $\phi_{p,e}$ decrease with $\xi$ (most rapidly around dark notch), reaching new asymptotic values at $\xi \to +\infty$, such that phase difference $\phi_e - \phi_p$ again equals to $\pi$ [Fig. 3(a)]. Total phase variation in each component does not exceed $-\pi$ in this case. In contrast, in anti-dark solitons phase grows in both photonic and excitonic components [Fig. 3(b)]. Phase variation in the photonic component always exceeds $\pi$, but remains below $2\pi$, while in the excitonic component phase variation exceeds $3\pi$ [in this case asymptotic value $(\phi_e - \phi_p)_{\xi \to +\infty} = 3\pi$]. Due to nontrivial phase distributions the amplitudes $a_{p,e}$ in dark solitons never reach zero values, except at the upper edge of the gap, where background vanishes. Thus, dark solitons in this system are in fact "gray".

Representative properties of dark solitons are summarized in Fig. 4. To quantify the degree of deformation of carrying plane wave induced by the soliton it is convenient to introduce renormalized energy flow $U$ and integral widths $w_{p,e}$ of the dip/hump in the soliton profile:

$$U = U_p + U_e = \int_{-\infty}^{\infty} [(a_p - a_{p\infty})^2 + (a_e - a_{e\infty})^2] d\xi ,$$
$$w_{p,e}^2 = 4U_{p,e}^{-1} \int_{-\infty}^{\infty} \xi^2 (a_{p,e} - a_{p,e\infty})^2 d\xi. \quad (7)$$

It is also instructive to introduce *total amplitudes* of photonic and excitonic components as $\gamma_{p,e} = \max(a_{p,e}) - \min(a_{p,e})$, where maximal and minimal values are defined over the entire $\xi$-axis. Renormalized energy flow of dark soliton is a nonmonotonic function of propagation constant $b$ [Fig. 4(a)]. It vanishes at the upper edge of the gap $b = b_+$ and also at larger propagation constant value $b = b_u$ determining upper border of finite existence domain of the solitons. At both edges of this domain the solitons transform into plane waves. Lower and upper limits of the existence domain are indicated on the plane $(v, b)$ in Fig. 1(c) by white lines with circles. The domain of existence is most extended at $v = 0.5$ and shrinks completely at $v \to 0, 1$. Analogously, largest $U$ values (largest plane wave deformation) is achieved at $v = 0.5$ in the depth of the existence domain [this also follows from nonmonotonic dependence of the total soliton amplitude on $b$ from Fig. 4(d)]. Total phase shift in photonic component acquired between $\xi = -\infty$ and $\xi = +\infty$ [Fig. 4(b)] approaches $-\pi$ at $b \to b_+$, indicating that it is in this regime, close to the upper gap edge, the soliton becomes "black", although amplitude of the background also vanishes in this limit. Phase shift decreases toward upper edge of the existence domain and vanishes at $b = b_u$. The soliton is narrowest in the depth of the existence domain, while close to its edges the integral widths of both components diverge [Fig. 4(c)].

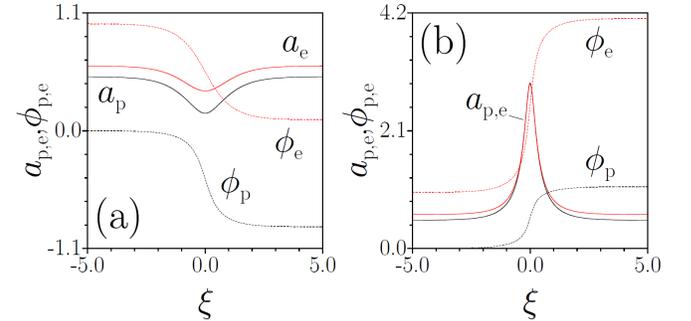

Fig. 3. (Color online) Field modulus (solid lines) and phase (dashed lines) distributions in dark (a) and bright (b) solitons at $b = 1.2$, $v = 0.5$. For illustrative purposes phases $\phi_{p,e}$ are divided by $\pi$.

The properties of anti-dark solitons substantially differ from those of dark states. Anti-dark solitons also exist within limited interval of propagation constants, adjacent to the upper edge of the forbidden gap. The upper edge of their existence domain nearly exactly coincides with that for their dark counterparts [see curve marked $b_u$ in Fig. 1(c)]. At the same time, the renormalized energy flow of the anti-dark solitons grows towards the upper edge of the gap [Fig. 5(a)], where such states turn into bright solitons on the zero background. While the background vanishes in this limit, the maximal soliton amplitude remains nonzero, as shown in Fig. 5(d). It should be mentioned that in this regime the profiles of photonic and excitonic components are very close. Close to the upper edge of the existence domain the phase distributions become extremely wide. Their width drastically exceeds that of the localized hump on top of the plane wave background. When the accumulated phase in the photonic component exceeds $2\pi$, it becomes impossible to find the anti-dark solitons with the constant phase on the tales [Fig. 5(b)]. The accumulated phase of the photonic component approaches $\pi$ at the lower edge of the existence domain. Note that delocalization of the phase distributions in the anti-dark solitons close to the upper edge of their existence domain is accompanied only by a slight increase of the integral pulse widths.

To study soliton stability we introduced small perturbations to the soliton profiles, substituted perturbed solutions into Eqs. (1), linearized them, and solved the corresponding linear eigenvalue problem. The outcome of stability analysis is that dark solitons are stable at $b_+ < b < b_{cr}$, in the part of their existence domain adjacent to the upper edge of the forbidden gap [see Fig. 1(c)]. An instability of dark solitons at $b > b_{cr}$ is associated exclusively with the instability of the *background plane wave*, and not with an instability of the soliton core. As a result, calculated growth rates $\delta$ for solitons coincide with modulation instability growth rates in Figs. 1(c),(d). Stability analysis for the anti-dark solitons reveals their instability in the entire existence domain. Their instability is associated with soliton core and leads to appearance of radiation, typically at the left soliton wing. Note, that dark solitons reported here can be excited dynamically with broad input pulses having properly engineered phase distributions.

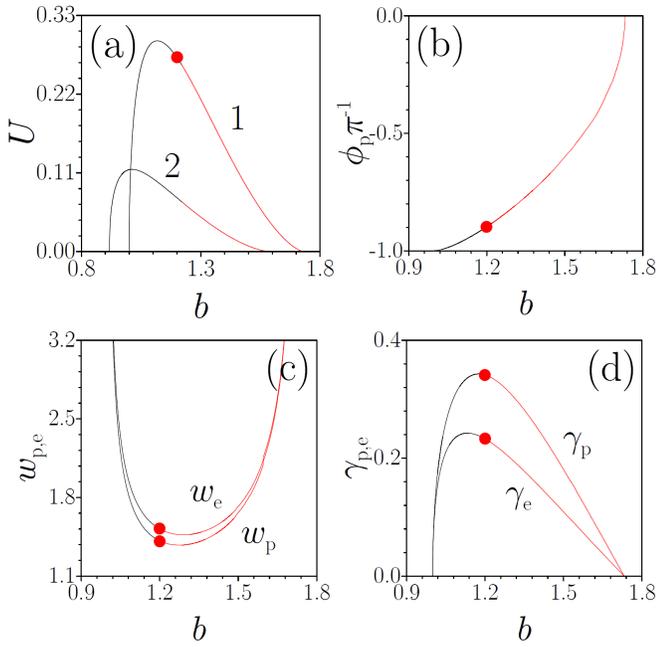

Fig. 4. (Color online) (a) Renormalized energy flow versus propagation constant for dark solitons at $v=0.5$ (curve 1) and $v=0.3$ (curve 2). Total phase shift for photonic component (b), widths $w_{p,e}$ (c), and amplitudes $\gamma_{p,e}$ (d) of photonic and excitonic components versus $b$ at $v=0.5$. Stable (unstable) branches are shown black (red). Red dots correspond to soliton from Fig. 3(a).

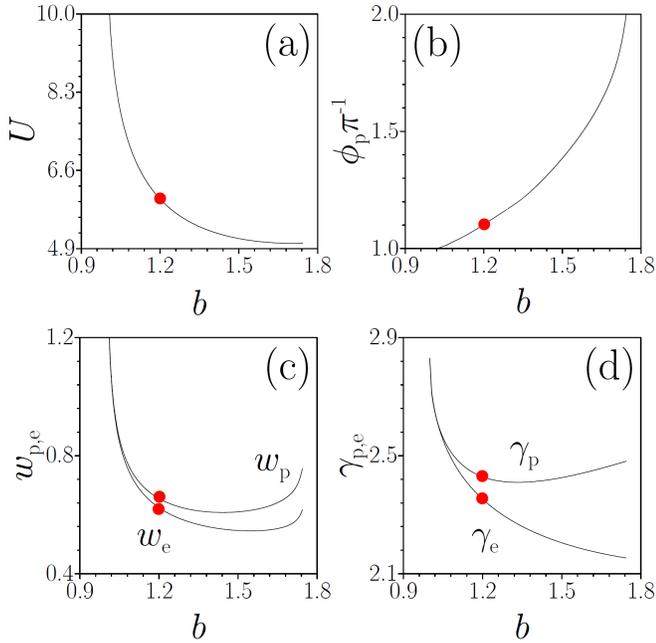

Fig. 5. (Color online) The same as in Fig. 4, but for anti-dark solitons. Red dots correspond to soliton from Fig. 3(b).

In Fig. 6 we show evolution of stable dark soliton in the presence of losses. Soliton propagates in a stable fashion adiabatically adjusting its parameters due to slow decrease of the background caused by losses.

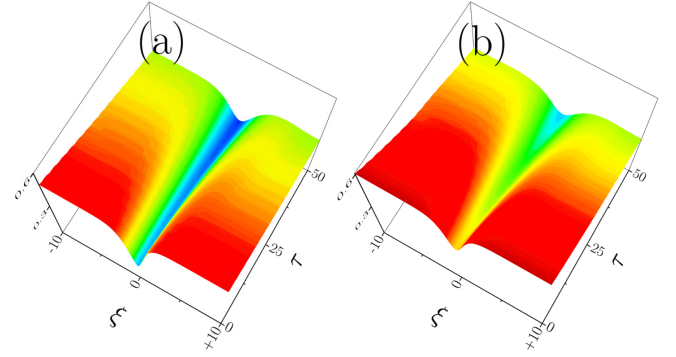

Fig. 6. (Color online) Evolution dynamics of dark soliton with $b=1.2$, $v=0.5$ in the presence of losses $\alpha_p = 0.0236$, $\alpha_e = 0.0040$. (a) Photonic $|\psi_p|$ and (b) excitonic $|\psi_e|$ components are shown.

Summarizing, we predicted that strong coupling between photons and excitons in waveguides with embedded quantum wells can result in formation of stable dark and unstable anti-dark light-matter solitons.

D.V.S. acknowledges support through the ITMO visiting professorship scheme. This work is supported by the Russian Federal Target Program (Project No. 14.587.21.0020), by the EU network project LIMACONA (Project No. 612600) and by the Leverhulme Trust Grant No. RPG-2012-481.